\documentclass[preprint,amssymb, aps, prl]{revtex4-1}
\usepackage{amsmath}
\usepackage{graphicx}

\begin{document}

\title{Relativistic electron beam acceleration in a zero-slippage \\ terahertz-driven Inverse Free Electron Laser scheme}

\author{E. Curry}
\email{ejcurry@physics.ucla.edu}
\affiliation{Department of Physics and Astronomy, UCLA, Los Angeles, California 90095, USA}

\author{S. Fabbri}
\affiliation{Department of Physics and Astronomy, UCLA, Los Angeles, California 90095, USA}

\author{J. Maxson}
\altaffiliation[Now at: ]{CLASSE, Cornell University, 161 Synchrotron Drive, Ithaca, New York 14853-8001, USA}
\affiliation{Department of Physics and Astronomy, UCLA, Los Angeles, California 90095, USA}

\author{P. Musumeci}
\affiliation{Department of Physics and Astronomy, UCLA, Los Angeles, California 90095, USA}

\author{A. Gover}
\affiliation{Faculty of Engineering, Department of Physical Electronics, Tel-Aviv University, Tel-Aviv 69978 Israel}

\begin{abstract}
THz radiation promises breakthrough advances in compact advanced accelerators due to the high frequency and GV/m fields achievable, orders of magnitude larger than in conventional radiofrequency (RF) based accelerators. Compared to laser-driven schemes, the large phase acceptances of THz-driven accelerators are advantageous for operation with sizable charge. Despite burgeoning research, THz sources, particularly laser-based ones, cannot yet compete with the efficiency of RF amplifiers for high average current accelerators. Nevertheless, THz-based phase space manipulation is of immediate interest for a variety of applications, including generation and diagnostics of ultrashort bunches for electron diffraction/microscopy and compact free-electron laser applications.

The challenge of maintaining overlap and synchronism between an electron beam and short laser-generated THz pulse has so far limited interactions to the few mm scale. We discuss a novel scheme for simultaneous group and phase velocity matching of nearly single-cycle THz radiation with a relativistic electron beam for meter-scale interaction. We demonstrate energy modulations of up to 150 keV using modest THz pulse energies ($\leq \,1 \,\mu$J). We apply this large and efficient energy exchange for beam compression and time-stamping of a relativistic beam, paving the way towards realizing the unique opportunities enabled by laser-based THz accelerators.
\end{abstract}

\maketitle
\date{today}

The continuous progress in the development of high peak and average power sources in the THz frequency range is closely followed by modern accelerator and beam physics, as THz waves have the potential to bridge the gap between radio frequency and laser frequencies in particle accelerator applications. THz-based acceleration, even though still at its infancy, retains key advantages from both regimes, namely the capability to sustain high-gradient fields and the broad phase acceptance window associated with millimeter-waves. Recent important proof-of-principle results have been obtained using corrugated waveguides and dielectric based structures. For example, using an iris-loaded waveguide structure, acceleration of 7 keV in 6 mm was reported by Nanni et al.~\cite{THzlinacNanni}. Similarly, THz waves have been successfully used to accelerate and manipulate electron beams at the injector \cite{THzgunHuang,THzgunFallahi, nanotipsRopers} and in non relativistic beamlines \cite{THzcontrolBaum,THzstreakFruhling}.

There are various techniques available to generate high power THz radiation. Beam-based sources are the most efficient and powerful \cite{FACETWu}. O'Shea et al. observed an energy gradient over 1 GeV/m using a beam-driven dielectric wakefield accelerator in the THz regime \cite{FACET:THzE210}. These sources are dependent upon the availability of intense high energy drive electron beams \cite{AWAsourceAntipov, SPARCChiadroni}. Laser-based THz sources \cite{THzgen10uJYeh,THzgenHebling} utilize ultrafast laser systems and access very high (up to GV/m) electric fields even with modest pulse energy due to the very short pulsewidth (near single-cycle) of the generated THz radiation. While beam-based sources might be the only viable solution for high efficiency applications such as high average power accelerators, the compactness, flexible temporal structure and inherent synchronization with an external laser signal make a laser-based THz source uniquely attractive for electron beam manipulation and diagnosis. Among these, optical rectification in non-linear crystals is particularly appealing, given the prospects for improving conversion efficiency and scaling to higher THz pulse energies \cite{improvingORFulop,coolingHuang,THzgenDASTVicario} and harnessing peak field enhancement \cite{THzfieldenhanceBagiante,Fabianska}.

One of the main drawbacks to beamline applications of laser-based THz sources is that the short duration of near-single cycle pulses strongly limits the temporal overlap between electron beam and high-intensity THz field \cite{THzIFELMoody, THzLPSJamison, THzLPSHebling}. In metallic or dielectric structures it is very difficult to simultaneously tune both the phase and group velocity of the radiation, as required in order to keep the particles overlapped with the envelope of the THz pulse, while riding at a constant accelerating or decelerating phase \cite{Wong:THzwaveguides,accelphaseWalsh}. The result is a group velocity mismatch which quickly degrades the efficiency of the energy exchange, posing a severe limitation on the length of the interaction (typically less than a few mm).

In \cite{backgroundCurry} we presented a THz-driven inverse free-electron laser (IFEL) scheme where the length of the interaction can be dramatically increased through a ``zero-slippage" interaction scheme using a curved parallel plate waveguide (CPPWG) to slow down the THz group velocity. The main advantages of this so called zero-slippage IFEL coupling scheme are the broad spectral bandwidth for resonant interaction and the large transverse acceptance due to the far-field nature of the undulator ponderomotive coupling. Furthermore, as the waveguide effectively counteracts the transverse spread of the THz pulse due to diffraction, it is natural to extend the length along which efficient energy exchange occurs to a meter-scale interaction.

\begin{figure*}[!htb]
\centering
\includegraphics[width=1.01\textwidth]{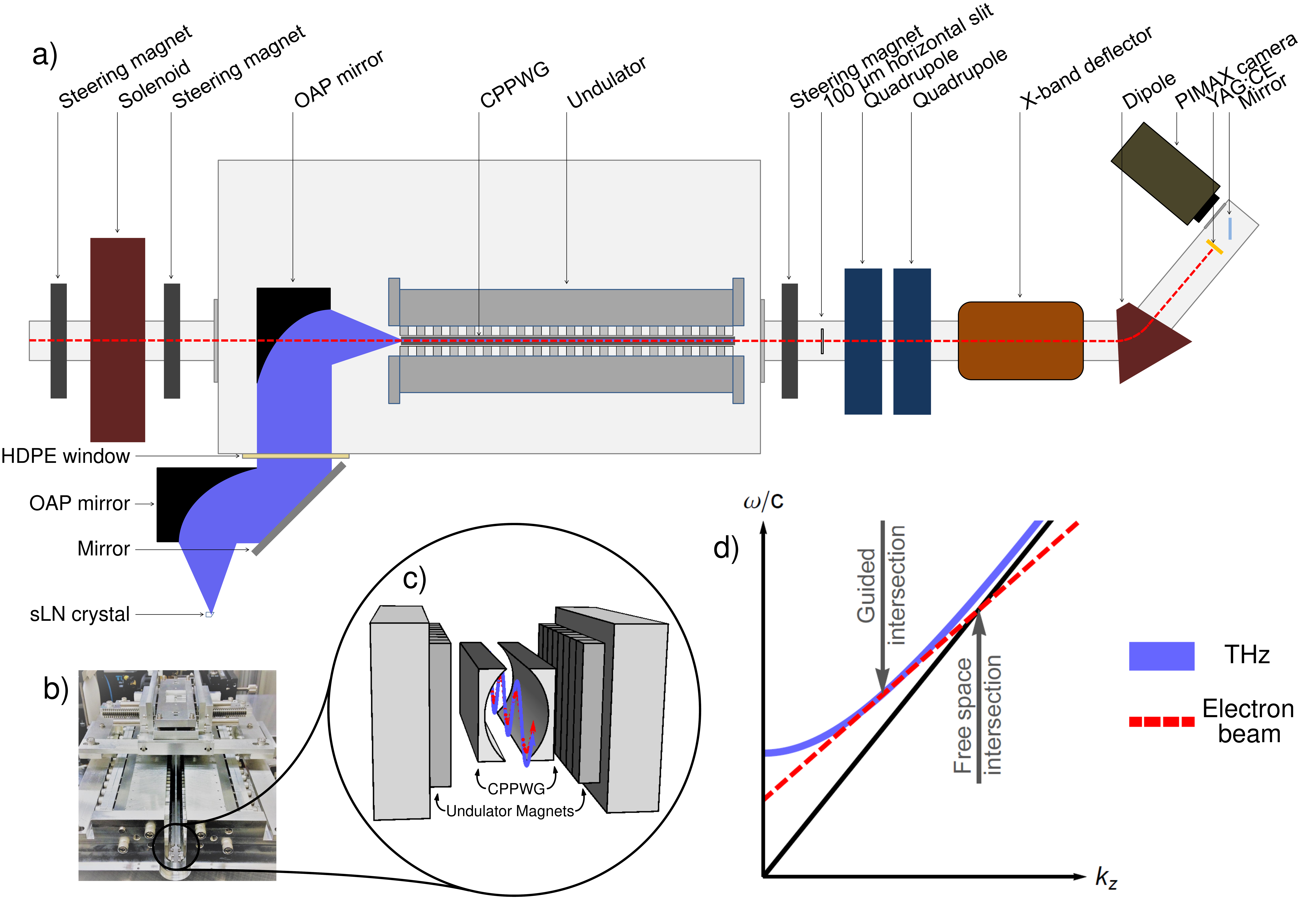}
\caption{\textbf{Zero-slippage THz-driven IFEL scheme.} a) Diagram of the experimental area on the PEGASUS beamline. The electrons are moving from left to right (dashed red). b) Photograph of undulator with nested CPPWG apparatus alongside c) a visualization of the IFEL interaction scheme. d) THz dispersion curves are shown for propagation in free space (black) and guiding by the CPPWG (light blue). Intersection with the dashed red line, given by Eq.~\eqref{phase_synchronism} for the electron beam, indicates IFEL phase synchronism. The tangential intersection possible in the guided case corresponds to group velocity matching of the electron beam and THz pulse for a ``zero-slippage" interaction.}
\label{setup_diagram}
\end{figure*}

In this paper we discuss the first experimental demonstration of this scheme (see Fig. \ref{setup_diagram} and Methods section for details). By copropagating a single cycle 1 $\mu$J THz pulse with a 4-9 MeV high brightness electron beam from an RF photoinjector \cite{RFgunAlesini} in a 30 cm long planar undulator and a tunable plate spacing CPPWG, we observed up to $150$ keV peak-to-peak induced energy modulation. Results from beam energy scans at different plate spacings are used to demonstrate the resonant nature of the interaction and validate the numerical models for this scheme \cite{CurryCuba}. The induced energy modulation was observed to scale linearly with the THz field amplitude as expected, opening the door to scaling this scheme to significantly higher energies. Longitudinal phase space (LPS) measurements of a multi-ps input electron beam reveal the full sinusoidal energy modulation due to the THz field and can be used to retrieve the relative time-of-arrival between electrons and radiation. An initial demonstration of the application of this scheme for THz-driven bunch length compression was also carried out resulting in a factor of two temporal compression.

\section{Results}

\textbf{Zero-slippage IFEL scheme.} The IFEL interaction in a planar undulator is governed by the phase synchronism condition, given in terms of the THz frequency $\omega$ and undulator wavenumber $k_u$ as
\begin{equation}
\frac{\omega}{c \bar{\beta_z}} = k_z(\omega)+k_u
\label{phase_synchronism}
\end{equation}
which, for a given undulator of period $2\pi/k_u$ and normalized magnetic field amplitude $K=\frac{eB_u}{k_u mc}$, can be met either by adjusting the normalized longitudinal electron velocity $\bar{\beta_z} = \sqrt{1-\frac{1+K^2/2}{\gamma^2}}$, or the THz wavenumber, $k_z(\omega)$, whose $\omega$ dependence is set by the dispersion relation of the waveguide. In practice, the resonance can be met by varying the beam energy or the CPPWG plate spacing. The extra degree of freedom can be used to match the THz group velocity with the electron propagation $\bar{\beta_z}$, resulting in a ``zero-slippage" interaction, wherein the electron bunch and THz pulse remain overlapped throughout the undulator \cite{backgroundCurry}.

\begin{table}
\begin{tabular}{lc}
\hline
\hline
Bunch energy & 4 - 9 MeV\\
Undulator period & 3 cm \\
Undulator parameter, K & 1.27 \\
\# of undulator periods & 10 \\
CPPWG spacing, b & 1.8-2.7 mm \\
Plate curvature radius & 2 mm \\
Peak frequency & 0.84 THz \\
Pulse energy & 1 $\mu$J \\
\hline
\hline
\end{tabular}
\caption{IFEL Experimental Parameters}\label{parameters}%
\end{table}

\begin{figure}[!htb]
\centering
\includegraphics*[width=1.05\textwidth]{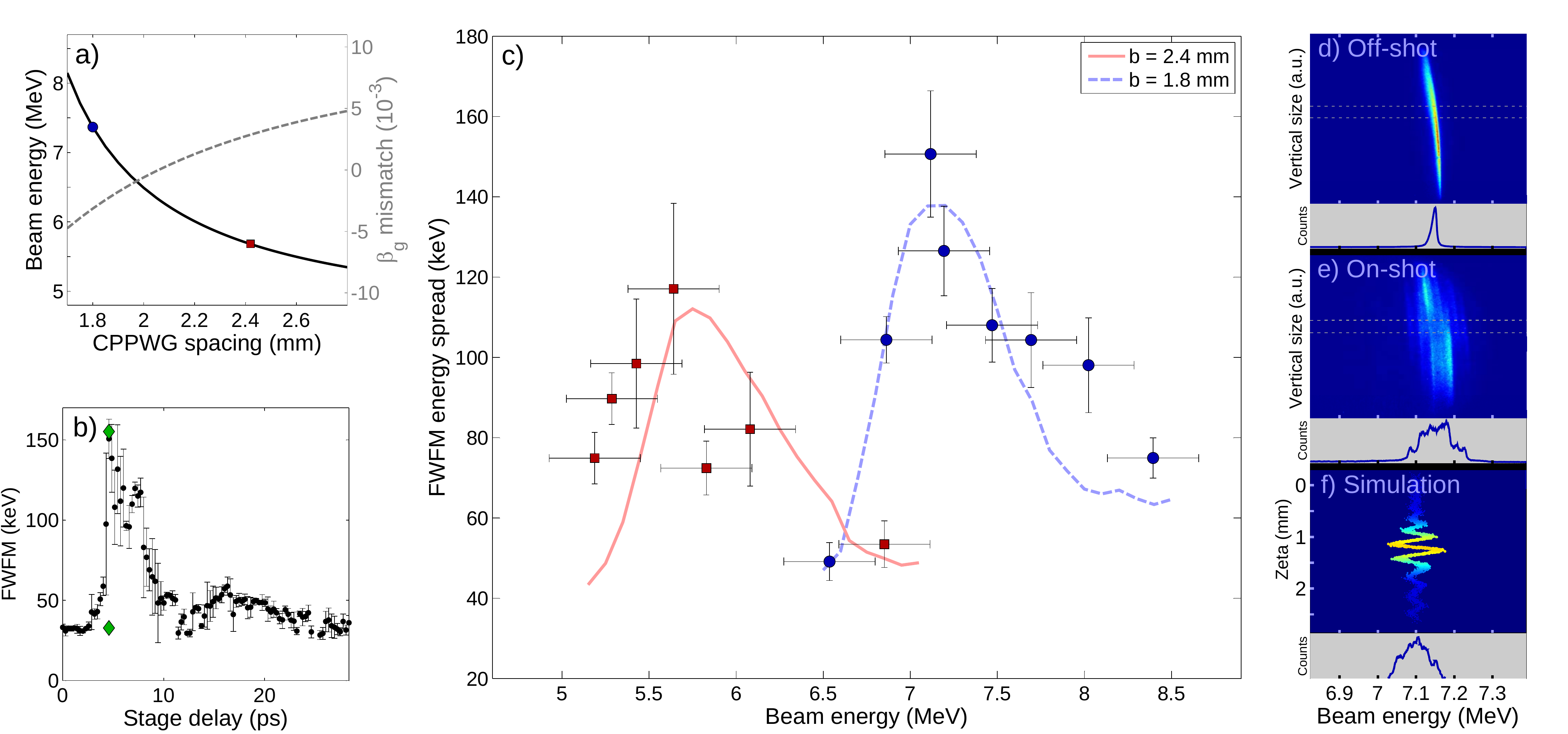}
\caption{\textbf{Resonant energy scans for IFEL interaction at different CPPWG spacings.} a) Plot showing the shift in resonant energy for a given CPPWG spacing, using the parameters given in Table~\ref{parameters}, for a single THz frequency. The points refer to spacings where measurements in plot c) were taken. The corresponding group velocity mismatch is given by the dashed gray line with scale indicated on the right most axis. b) Plot of beam energy spread as a function of THz arrival time, with two green diamonds corresponding to the THz off and on-shot images in d) and e). 
c) Scan of the maximum induced energy spread vs beam energy for two different plate spacings. Each point is an average of on-shots from the stage position corresponding to peak energy modulation. The dashed lines indicate the FWFM energy modulation from simulation. Consecutive raw beam images, with d) THz off and e) THz on, showing the change in energy spread after IFEL interaction and f) simulation of electron beam phase space distribution after IFEL interaction. The corresponding energy distribution is shown at the bottom of each plot.}
\label{energy_scan}
\end{figure}

In Fig.~\ref{setup_diagram}d, the dispersion curve of the electron beam in the undulator, as determined by Eq.~\eqref{phase_synchronism}, is plotted alongside the THz dispersion curves for the case of free space propagation and CPPWG guiding. In free space, the curves intersect and phase matching is satisfied only for one frequency, and thus no group velocity matching; when the waveguide is used, the THz and electron beam curves can be tangent to each other indicating group velocity matching and broadband interaction (the zero-slippage condition). The beam energies that satisfy the phase synchronicity condition (i.e. Eq. \ref{phase_synchronism}) for different waveguide spacings are shown in Fig.~\ref{energy_scan}a. The undulator and radiation parameters used for this plot correspond to the experimental ones which are summarized in Table \ref{parameters}. Notice how increasing the waveguide spacing results in a smaller resonant energy, since while the group velocity increases, the phase velocity decreases. The difference between average longitudinal beam velocity and THz group velocity is also shown by the dashed gray line. As discussed below, dispersion of the THz pulse in the waveguide as well as coupling efficiency affect the strength of the interaction for different waveguide spacings.

\textbf{THz interaction measurements.} To determine the optimal THz pulse timing, the path length of the THz source was adjusted with a delay stage while measuring the induced energy spread, see Fig.~\ref{energy_scan}b. The duration of the interaction is a convolution of the electron and THz bunch lengths. Two consecutive beam images displaying the electron energy spectrum for a THz off-shot and on-shot, from the peak of this timing scan are shown in Figure \ref{energy_scan}d and \ref{energy_scan}e, respectively. The induced energy spread is quantified by the full width full max (FWFM) of a small vertical slice in the image, where the FWFM is defined here as the width of the energy distribution at $10\%$ of the peak value. 

These measurements were taken using an electron beam with 1.5 ps rms bunch length, covering multiple periods of the THz waveform, which increases the window of overlap between the electron bunch and THz interaction region, therefore in each shot some electrons are accelerated and some decelerated. The energy distribution of the THz on-shot in Fig.~\ref{energy_scan}e is plotted at the bottom of the figure and it is symmetrically widened with respect to the initial distribution (see the corresponding sliced projection in Fig.~\ref{energy_scan}d). This distribution is expected since electrons are injected over all interaction phases and are equally likely to lose or gain energy. The full width of the final spectrum can be as large as $150$ keV in good agreement with our simulation model (discussed in detail below). In order to estimate the maximum energy gain (or loss) for single particles this value should be deconvolved with the initial distribution (which has a full width of 30 keV) and divided by two. Note that residual dispersion in the CPPWG stretches the broadband THz pulse to a few-cycle pulse, with a varying field amplitude determined by the overall pulse envelope. The discontinuous steps visible in the projected energy distribution of the THz on-shot are the result of the multicycle nature of the energy modulation, with portions of the beam undergoing different modulation amplitudes. The same feature is reproduced in the longitudinal phase space after a simulated beam interaction (shown in Fig.~\ref{energy_scan}f with the corresponding energy distribution).

To demonstrate waveguide-based tuning of the IFEL resonance condition, we performed measurements over a range of beam energies for different waveguide spacings (1.8 mm and 2.4 mm) which are shown in Fig.~\ref{energy_scan}c. Each point represents the maximum energy spread determined after scanning the THz arrival time at a particular beam energy. For smaller plate spacing, the resonant energy increases, consistent with our simple model of the interaction.

A simulation code which takes into account the different coupling efficiency as well as dispersion in the CPPWG was developed to understand quantitatively the strength of the beam-radiation interaction. Coupling efficiency is calculated using the transverse profile overlap integral of the THz free-space propagating and waveguide modes. The radiation field is evolved in the frequency-domain using the known waveguide dispersion. The code tracks the particles through the undulator using the input THz field as determined via electro-optic sampling (EOS) measurements and the initial beam energy spread as determined from spectrometer images with the THz source off. The beam charge in the experiment was kept below 1 pC so that the evolution of the radiation field is negligible. The simulation results reported in Fig.~\ref{energy_scan}c assume $25 \%$ power coupling of the 1 $\mu$J THz pulse, as measured at the sLN crystal and are in good agreement with the data. Residual differences are ascribed to experimental uncertainties related to the coupling of the THz pulse into the guide, misalignment of the CPPWG plates, and alignment of the electron beam through the guide and undulator axis. Note that additional interaction with higher order modes in the waveguide might further contribute to particle acceleration or deceleration, even though this effect is calculated to be responsible for less than 10 $\%$ of the final induced energy spread.

\begin{figure}[!htb]
\centering
\includegraphics*[width=1\textwidth]{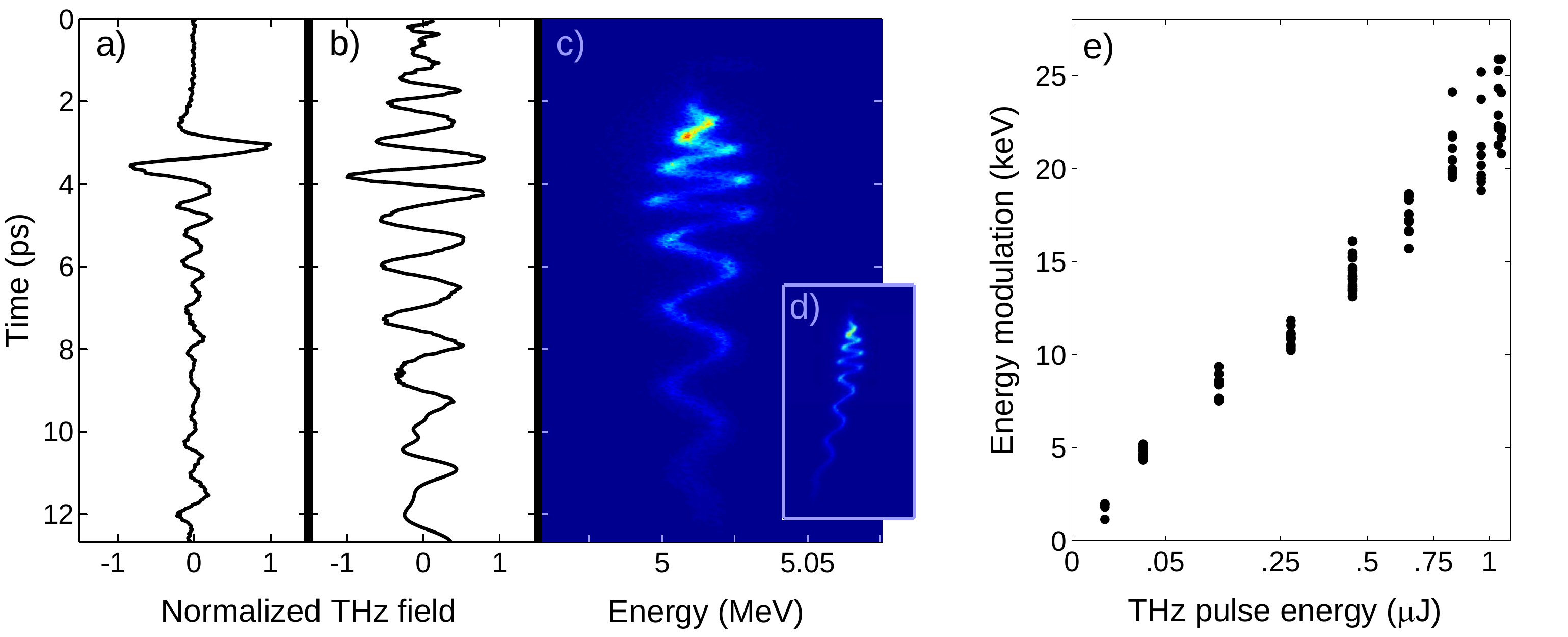}
\caption{\textbf{Longitudinal phase space of the electron beam.} a) Initial THz waveform determined via EOS measurements. b) THz pulse after a 30 cm CPPWG, using the dispersion relation to evolve the initial EOS trace. c) Longitudinal phase space measurement showing beam energy on the x-axis and temporal position on the y-axis with d) the corresponding raw beam image. e) Measurements of the maximum peak-to-peak energy modulation from LPS images as a function of THz energy as measured by the pyrometer. Using a quadratic x-axis scale, the linear relationship between THz field and energy modulation is evident. These measurements were taken with 5 MeV beam energy and 2.7 mm waveguide spacing.}
\label{EOS_scaling}
\end{figure}

\textbf{Longitudinal phase-space measurements.}
For a more complete picture of the IFEL interaction between the THz pulse and electron beam, we analyzed images of the longitudinal phase space (LPS) of the beam, wherein we map the energy gain vs bunch temporal coordinate. After the IFEL, the beam was deflected vertically in an X-band cavity before undergoing the energy-dependent deflection in the dipole magnet \cite{deflectorMaxson}. A slit was used to limit the Panfosky-Wenzel deflector-induced energy spread to $< 10$ keV and two quadrupoles minimized the betatron beam sizes on the screen \cite{pegasusLPSMoody}. Figure \ref{EOS_scaling}d shows an example of a raw LPS image.

The LPS measurements reveal the multi-cycle energy modulation that we expected from interaction with the dispersed THz pulse. In the zero-slippage scheme, the overall energy modulation reflects the waveform of the evolving THz pulse envelope after pulse dispersion, see Fig.~\ref{EOS_scaling}b.

The slanted LPS distribution in the beam image in Figure \ref{EOS_scaling}d shows the underlying energy chirp caused by RF curvature during the initial acceleration of the beam which can be subtracted off from the THz on-shot distribution in order to quantify the effect of the IFEL interaction (see Figure \ref{EOS_scaling}c). A direct measurement of the beam time-of-arrival jitter with respect to the THz can be obtained by monitoring the position of the energy modulation peaks with respect to the electron beam head and yields 400 fs rms temporal jitter in agreement with previous measurements on the Pegasus beamline \cite{GeSwitchCesar}. The peak-to-peak energy modulation can be directly measured from these images. A scan in the THz energy controlled by varying the amount of infrared laser illuminating the sLN crystal displays the expected linear relation between the maximum energy modulation and the THz field amplitude (see Figure \ref{EOS_scaling}e). Note that the data here corresponds to a larger than optimal waveguide spacing and low incident THz field so that the interaction was not maximized. The reported value is the maximum peak-to-peak energy spread from a moving average of the pixel distribution, after filtering to remove background and transforming the original image (see Figure \ref{EOS_scaling}d) to remove RF curvature.


\textbf{THz-driven bunch length compression} One attractive application of this interaction is the possibility of inducing a positive energy correlation on a short electron beam injected at the zero-crossing of the wave. A dispersive section (which at the moderate beam energies used here can be just a meter-scale drift) will cause the faster particles in the tail to catch up with the head of the beam, shortening the bunch length. Due to the high frequency of the THz wave, the benefits of a THz-based bunch compressor include both the possibility of inducing a steep energy chirp in the beam longitudinal phase space and reducing the timing jitter with respect to the laser pulse which is used for THz generation \cite{backgroundCurry}.

\begin{figure}[!htb]
\centering
\includegraphics[width=.7\textwidth]{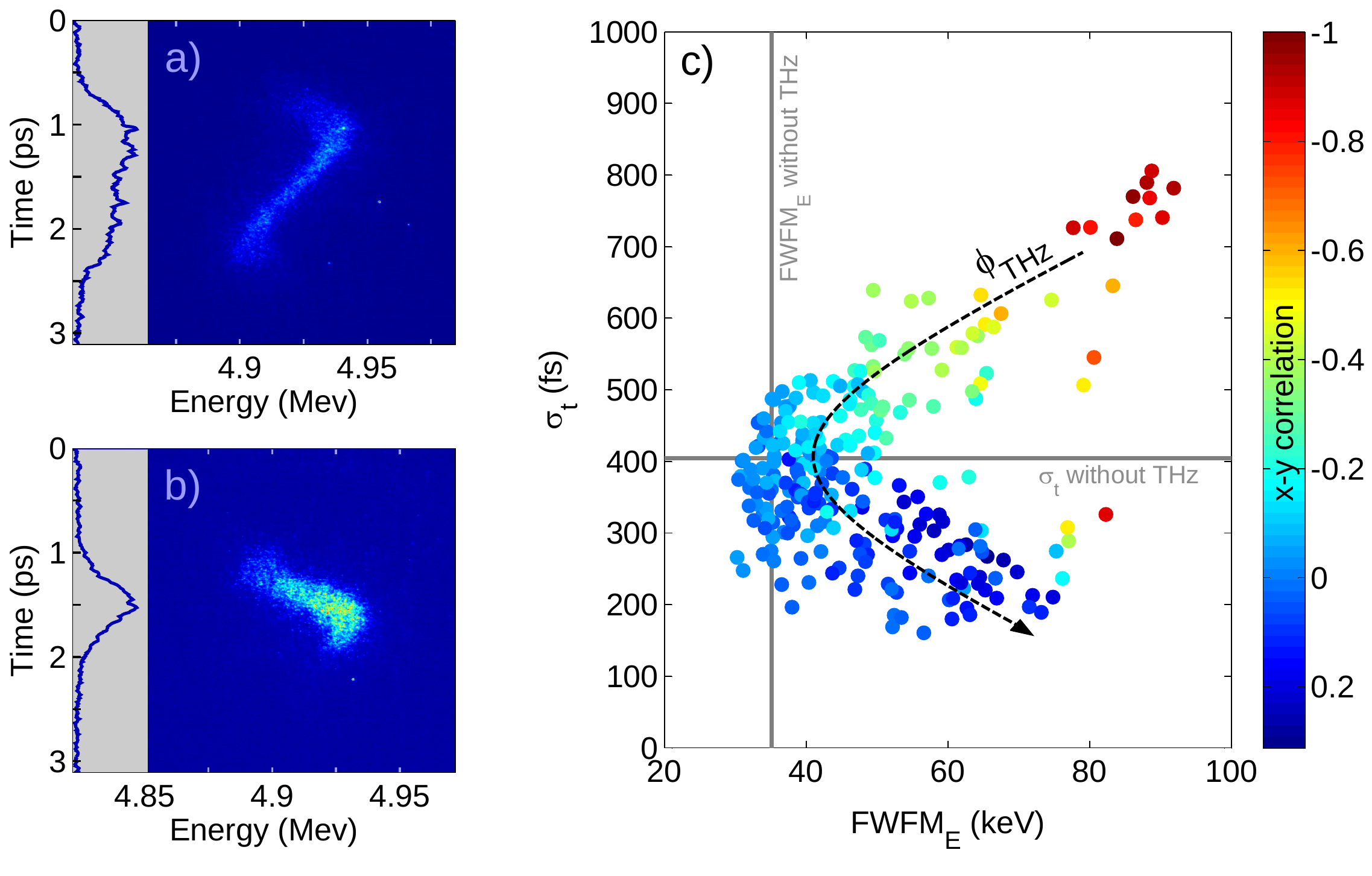}
\caption{\textbf{Electron bunch length compression.} a) and b) are consecutive LPS images, showing a case of positive energy chirp resulting in decompression and negative energy chirp resulting in compression, respectively. c) A plot of rms bunchlength vs. energy spread determined for many LPS images. Color is determined by the energy-position correlation in the LPS image, with cool colors corresponding to compression and warm colors corresponding to decompression of the beam. Gray lines show the mean energy spread and bunch length of the beam without THz interaction. An arrow indicates the transition from decompression to compression.}
\label{bunching}
\end{figure}

In our experiment we re-compressed the photocathode driver laser to a 100 fs pulse length in order to produce an electron beam short enough to fit in the linear part of the THz wave and test this concept for THz-driven bunch compression. The electron bunch length was measured by an RF streak camera located 1 m downstream of the undulator. In practice, because of relatively large ($~$400 fs rms) time of arrival jitter of the electron beam at PEGASUS associated with amplitude and phase fluctuations in the aging RF system, the injection phase in the THz wave could not be precisely controlled, resulting in sampling of different regions of the THz waveform with each shot. This translated into a corresponding range of induced energy chirps which, depending on the sign of the energy correlation, led to compression or decompression of the beam in the subsequent drift. For cases where the beam is injected at a phase with negative field gradient (so that the head is accelerated to higher energy than the tail), the final bunch LPS is elongated, with a negative energy-position correlation and significant energy spread, as in Fig.~\ref{bunching}a.

Full compression can be obtained when the dispersion in the drift following the undulator is sufficient to allow the back of the bunch to catch the head. For the measurements discussed here, the undulator-streak camera distance was limited to 1 m and slightly shorter than what was required to obtain full compression (for the maximum induced energy modulation). Even at the shortest bunch lengths, the LPS measurements still show a nonzero energy-position correlation, as in Fig.~\ref{bunching}b. In the absence of interaction, the electron beam from the linac was measured to be 400 fs long, too large to fit fully within a half period of the THz waveform, resulting in the LPS aberrations easily noticeable in the distributions in Fig.~\ref{bunching}a and \ref{bunching}b.

Figure \ref{bunching}c shows a plot of bunch length vs. energy spread for a set of short beam measurements. The color indicates the energy-position correlation in the beam, or equivalently, the phase of the THz waveform with which the beam interacted. There is a clear transition from interactions that are strongly decompressing, indicated by a negative energy-position correlation shown in warm colors, to interactions that are compressing the beam, indicated by a positive energy-position correlation shown in cool colors. As expected, the decompressed beam has large energy spread and bunch length. Where the beam interacts with a small THz field gradient, there is minimal energy chirp, i.e. near-zero energy-position correlation, and the energy spread and bunch length are largely unchanged. At the bottom of the plot, the energy spread increases, while the bunch length decreases to a minimum of 200 fs, demonstrating THz-driven bunch compression.

The results of the zero-slippage IFEL coupling scheme presented here show a robust and flexible technique for THz-driven beam manipulation. The performance of the IFEL produces a significant energy modulation, up to 150 keV, of the relativistic electron beam. The primary factor limiting the energy modulation in the results presented here is the initial THz pulse energy. THz generation techniques are improving rapidly, and, already, pulse energies of orders of magnitude larger are readily available for THz-driven beam manipulation\cite{THzlinacNanni}. We demonstrated that the small acceleration gradient, 0.25 MeV/m, achieved here for relativistic electrons scales linearly with the initial THz field available. An increase of more than two orders of magnitude would then be possible when using higher power laser-based THz generation sources. Because of the group-velocity matching properties of the zero-slippage scheme, the energy modulation could also be increased by further extending the undulator and CPPWG length. The limit in this case is set by the dispersive pulse broadening in the guide and attenuation losses which reduce the THz field gradient available for chirping the beam.

The success of this first demonstration of a zero-slippage IFEL for THz-driven beam manipulation is bolstered by the versatility and accessibility of the design. Further applications of this THz coupling scheme include transverse deflection for longitudinal beam diagnostic \cite{CurryCuba} (using an odd-symmetry mode in the waveguide) and ultimately THz broadband amplification when a high charge beam is decelerated in a tapered undulator \cite{Duris:TESSA}.

\section*{Acknowledgments}
This work has been supported by DOE grant DE-FG02-92ER40693 and NSF grant PHY-1415583 and PHY-1734215 and partial support by the US-Israel Binational Science Foundation (BSF). The authors acknowledge useful discussions with G. Andonian and D. Cesar.

\section*{Author contributions}
E.C. and P.M. conceived and designed the experiment. A.G. provided theoretical support. S.F. provided technical support for the undulator and waveguide assembly and construction. J. M. contributed to the electron beamline diagnostics and participated in the data analysis. The manuscript was prepared by E.C. and P.M. with contributions from all coauthors.

\section{Methods}
\textbf{THz pulse generation and characterization.}
The THz pulse used in these experiments was generated in a stoichiometric Lithium Niobate (sLN) crystal using a pulse-front-tilt scheme. Up to 1 mJ of 800 nm laser was used in the optical rectification process with a maximum conversion efficiency of 0.1 $\%$. The THz radiation generated at the crystal was then collected by an off-axis parabolic mirror (OAP) and brought into the vacuum through a high transmission ($>$ 80 $\%$) HDPE window. A 15 cm focal length OAP with a 2 mm hole for electron transmission was used to focus the THz at the entrance of the tunable spacing of the CPPWG ($b$ = 1.5-3 mm). Electro-optic sampling in a ZnTe crystal was used to measure the temporal profile of the THz pulse. A pyrometer was used to measure THz pulse energies up to 1 $\mu$J, which depending on the coupling efficiency into the waveguide (itself a function of $b$), yields an estimate of 4-8 MV/m for the maximum THz field of the $TE_{01}$ mode of the CPPWG.

\textbf{RF photoinjector beamline.}
The 4-9 MeV beam used in this experiment was generated by illuminating a Cu cathode in a high field 1.6 cell S band gun. The beam energy was tuned using a booster linac cavity which also allowed to control the final energy spread. With the THz source off, the initial beam energy spread was minimized to 30 keV FWFM by adjusting the RF gun and linac phase. A solenoid was used to focus the electron beam from the Pegasus injector through the small hole in the THz-focusing OAP to a 200 $\mu$m waist at the entrance of the undulator.

Downstream of the undulator, a dipole magnet is used to horizontally disperse the energy distribution after the interaction on the spectrometer screen and a 9.6 GHz transverse deflecting cavity can be used to streak the longitudinal beam profile onto the vertical axis. A 100 $\mu$m horizontal slit and two quadrupoles before the cavity are used to maximize the resolution in the LPS measurements. A fluorescent high yield screen and intensified camera were used to detect the particles on the spectrometer. 

\textbf{Planar undulator.}
The planar undulator for this experiment was designed to provide resonant coupling between the electron energies readily available at the UCLA Pegasus injector and the existing THz source. A Halbach style undulator magnet of total length 30 cm having 10 periods of length 3 cm and gap 14 mm was built. The magnetic field amplitude was measured to be $B = 0.45$ T corresponding to a normalized vector potential $K$ = 1.27.  Special entrance and exit sections were tailored to keep the wiggling trajectory on axis. The final on-axis magnetic field profile, measured by Hall probe scans, was tuned by fine adjustments to the gap between individual magnets, based on the Radia 3d simulation model \cite{RadiaChubar}.

\textbf{Curved parallel plate waveguide.}
The CPPWG was suspended between the magnet arrays using a custom 6-axis self-centering mount with a spacing adjustment knob accessible from above. All components were designed for operation in vacuum. The variable jaw was used to control the spacing of the waveguide and ensure parallelism of the two halves. A camera located outside the vacuum chamber was focused at the entrance of the waveguide and used to determine the spacing between the plates with 100 $\mu$m accuracy.

\bibliographystyle{unsrt}

\begin{thebibliography}{10}

\bibitem{THzlinacNanni}
E.~A. Nanni, W.~R. Huang, K.~Ravi, A.~Fallahi, G.~Moriena, R.~J. Miller, and
  F.X. K{\"a}rtner.
\newblock{Terahertz-driven linear electron acceleration}.
\newblock {\em Nature communications}, 6, 2015.

\bibitem{THzgunHuang}
W.~R.~Huang, A.~Fallahi, X.~Wu, H.~Cankaya, A.-L. Calendron, K.~Ravi, D.~ Zhang, E.~A.~Nanni, K.-H.~Hong, and F.~X.~K{\"a}rtner.
\newblock{ Terahertz-driven, all-optical electron gun}.
\newblock {\em Optica}, 3(11):1209--1212, 2016.

\bibitem{THzgunFallahi}
A.~Fallahi, M.~Fakhari, A.~Yahaghi, M.~Arrieta, and F.~X.~K{\"a}rtner.
\newblock{ Short electron bunch generation using single-cycle ultrafast electron guns}.
\newblock {\em Physical Review Accelerators and Beams}, 19(8):081302, 2016.

\bibitem{nanotipsRopers}
L.~Wimmer, G.~Herink, D.~R.~Solli, S.~V.~Yalunin, K.~E.~Echternkamp, and C.~Ropers.
\newblock{Terahertz control of nanotip photoemission}.
\newblock {\em Nature Physics}, 10(6):432, 2014.

\bibitem{THzcontrolBaum}
C.~Kealhofer, W.~Schneider, D.~Ehberger, A.~Ryabov, F.~Krausz, and P.~Baum.
\newblock{All-optical control and metrology of electron pulses}.
\newblock {\em Science}, 352(6284):429--433, 2016.

\bibitem{THzstreakFruhling}
U.~Fr{\"u}hling, M.~Wieland, M.~Gensch, T.~Gebert, B.~Sch{\"u}tte, M.~Krikunova, R.~Kalms, F.~Budzyn, O.~Grimm, J.~Rossbach, and others.
\newblock{Single-shot terahertz-field-driven X-ray streak camera}.
\newblock{\em Nature Photonics}, 3(9):523--528,2009.

\bibitem{FACETWu}
Z.~Wu, A.~S.~Fisher, M.~C.~Hoffmann, S.~Bonetti, D.~Higley,
  and H.~Durr.
\newblock{THz light source at SLAC FACET user facility}.
\newblock In {\em Infrared, Millimeter, and Terahertz waves (IRMMW-THz), 2014
  39th International Conference on}, pages 1--2. IEEE, 2014.

\bibitem{FACET:THzE210}
B.~D.~O’Shea, G.~Andonian, S.~K.~Barber, K.~L.~Fitzmorris, S.~Hakimi, J.~Harrison, P.~D.~Hoang, M.~J.~Hogan, B.~Naranjo, O.~B.~Williams, V.~Yakimenko and J.~B.~Rosenzweig
\newblock{Observation of acceleration and deceleration in gigaelectron-volt-per-metre gradient dielectric wakefield accelerators}.
\newblock{\em Nature Communications} 7: 12763 (2016)

\bibitem{AWAsourceAntipov}
S.~Antipov, M.~Babzien, C.~Jing, M.~Fedurin, W.~Gai, A.~Kanareykin, K.~Kusche, V.~Yakimenko, and A.~Zholents.
\newblock{Subpicosecond bunch train production for a tunable mJ level THz source}.
\newblock {\em Physical review letters}, 111(13):134802, 2013.

\bibitem{SPARCChiadroni}
F.~Giorgianni, and others
\newblock{Strong nonlinear terahertz response induced by Dirac surface states in $Bi_2Se_3$ topological insulator}.
\newblock{\em Nature Communications} 7: 11421 (2016)

\bibitem{THzgen10uJYeh}
K.-L.~Yeh, M.~C.~Hoffmann, J.~Hebling, and K.~A.~Nelson.
\newblock{Generation of 10 $\mu$J ultrashort terahertz pulses by optical rectification}.
\newblock {\em Applied Physics Letters}, 90(17):171121, 2007.

\bibitem{THzgenHebling}
J.~Hebling, K.-L.~Yeh, M.~C.~Hoffmann, and K.~A.~Nelson.
\newblock{High-power THz generation, THz nonlinear optics, and THz nonlinear spectroscopy}.
\newblock {\em IEEE Journal of Selected Topics in Quantum Electronics},
  14(2):345--353, 2008.

\bibitem{improvingORFulop}
J.~A. F{\"u}l{\"o}p, L.~P{\'a}lfalvi, M.~C.~Hoffmann, and J.~Hebling.
\newblock{Towards generation of mJ-level ultrashort THz pulses by optical rectification}.
\newblock {\em Optics express}, 19(16):15090--15097, 2011.

\bibitem{coolingHuang}
S.-W.~Huang, E.~Granados, W.~R.~Huang, K.-H.~Hong, L.~E.~Zapata, and F.~X.~K{\"a}rtner.
\newblock{High conversion efficiency, high energy terahertz pulses by optical rectification in cryogenically cooled lithim niobate}.
\newblock {\em Optics letters}, 38(5):796--798, 2013.

\bibitem{THzgenDASTVicario}
C.~Vicario, B.~Monoszlai, and C.~P.~Hauri.
\newblock{GV/m single-cycle terahertz fields from a laser-driven large-size partitioned organic crystal}.
\newblock{\em Physical Review Letters}, 112(21):213901, 2014.

\bibitem{PPLNgenTHzJolly}
S.~W.~Jolly, F.~Ahr, N.~H.~Matlis, S.~Carbajo, K.~Ravi, T.~Kroh, J.~Schulte, D.~N.~Schimpf, A.~R.~Maier, F.~X.~K{\"a}rtner.
\newblock{Narrowband Terahertz Generation with Broadband Chirped Pulse Trains in Periodically Poled Lithium Niobate}.
\newblock{\em CLEO: QELS\_Fundamental Science}, FW4D--4, 2017.

\bibitem{THzfieldenhanceBagiante}
S.~Bagiante, F.~Enderli, J.~Fabia{\'n}ska, H.~Sigg, and T.~Feurer
\newblock{Giant electric field enhancement in split ring resonators featuring nanometer-sized gaps}.
\newblock{\em Scientific reports}, 5, 2015.

\bibitem{Fabianska}
J.~Fabia{\'n}ska, G.~Kassier and T.~Feurer.
\newblock{Split ring resonator based THz-driven electron streak camera featuring femtosecond resolution}.
\newblock{\em Scientific Reports} 4: 5645 (2014)

\bibitem{THzIFELMoody}
J.~T.~Moody, R.~K.~Li, P.~Musumeci, C.~M.~Scoby, H.~To, R.~Zgadzaj, E.~Gaul, and M.~C.~Downer.
\newblock{Longitudinal phase space manipulation of an ultrashort electron beam via THz IFEL interaction}.
\newblock In {\em AIP Conference Proceedings}, volume 1507, pages 722--727. AIP, 2012.

\bibitem{THzLPSJamison}
S.~P.~Jamison, T.~Thakker, B.~Muratori, Y.~M.~Saveliev, R.~J.~Smith,
 M.~Cliffe, W.~R.~Flavell, D.~M.~Graham, D.~J.~Holder, D.~Newton, and
 A.~Wolski.
\newblock{Phase space manipulation with laser-generated terhaertz pulses}.
\newblock Joint Accelerator Conferences Website, 6, 523--526, 2013.

\bibitem{THzLPSHebling}
J.~Hebling, J.~A.~F{\"u}l{\"o}p, M.~I.~Mechler, L.~P{\'a}lfalvi, C.~T{\H{o}}ke, and
 G.~Alm{\'a}si.
\newblock{Optical manipulation of relativistic electron beams using THz pulses}.
\newblock {\em arXiv preprint arXiv:1109.6852}, 2011.

\bibitem{Wong:THzwaveguides}
L.~J.~Wong, A.~Fallahi and F.~X.~K{\"a}rtner. 
\newblock{Compact electron acceleration and bunch compression in THz waveguides}.
\newblock{\em Optics Express} 21, 9792–9806 (2013).

\bibitem{accelphaseWalsh}
D.~A.~Walsh, D.~S.~Lake, E.~W.~Snedden, M.~J.~Cliffe, D.~M.~Graham, and S.~P.~Jamison.
\newblock{Demonstration of sub-luminal propagation of single-cycle terahertz pulses for particle acceleration}.
\newblock {\em Nature Communications}, 8(1):421, 2017.

\bibitem{backgroundCurry}
E.~Curry, S.~Fabbri, P.~Musumeci, and A.~Gover.
\newblock{THz-driven zero-slippage IFEL scheme for phase space manipulation}.
\newblock {\em New Journal of Physics}, 18(11):113045, 2016.

\bibitem{RFgunAlesini}
D.~Alesini, and others
\newblock{New technology based on clamping for high gradient radio frequency photogun}.
\newblock {\em Phys. Rev. STAB}, 18, 092001 (2015).

\bibitem{CurryCuba}
E.~Curry, S.~Fabbri, P.~Musumeci, and A.~Gover.
\newblock{Simulation of 3-D effects in THz-based phase space manipulation}.
\newblock {\em Nuclear Instruments and Methods in Physics Research Section A: Accelerators, Spectrometers, Detectors and Associated Equipment}, 2017.

\bibitem{deflectorMaxson}
J.~Maxson, D.~Cesar, G.~Calmasini, A.~Ody, P.~Musumeci, and D.~Alesini.
\newblock{Direct measurement of sub-10 fs relativistic electron beams with ultralow emittance}.
\newblock{\em Physical Review Letters}, 118(15):154802, 2017.

\bibitem{pegasusLPSMoody}
J.~T.~Moody, P.~Musumeci, M.~S.~Gutierrez, J.~B.~Rosenzweig, and C.~M.~Scoby.
\newblock{Longitudinal phase space characterization of the blow-out regime of rf photoinjector operation}.
\newblock {\em Physical Review Special Topics-Accelerators and Beams},
  12:070704, 2009.

\bibitem{GeSwitchCesar}
D.~B.~Cesar, P.~Musumeci, and D.~Alesini.
\newblock{Ultrafast gating of a mid-infrared laser pulse by a sub-pC relativistic electron beam}.
\newblock {\em Journal of Applied Physics}, 118(23):234506, 2015.

\bibitem{RadiaChubar}
O.~Chubar, P.~Elleaume, and J.~Chavanne.
\newblock{A three-dimensional magnetostatics computer code for insertion devices}.
\newblock {\em Journal of synchrotron radiation}, 5(3):481--484, 1998.

\bibitem{Duris:TESSA}
J.~P.~Duris, A.~Murokh and P.~Musumeci.
\newblock{Tapering enhanced stimulated superradiant amplification}.
\newblock{\em New Journal of Physics}, 17:063036 (2015)

\end{thebibliography}

\end{document}